\documentclass[prb,twocolumn,amssymb,letter]{revtex4}

\usepackage{bm}
\usepackage{graphicx}
\usepackage{dcolumn}
\usepackage[spanish]{babel}
\begin{document}

\title{Dispersi\'on inel\'astica de la luz por excitaciones electr\'onicas en \'atomos artificiales}
\author{Alain Delgado}
\affiliation{Centro de Aplicaciones Tecnol\'ogicas y
 Desarrollo Nuclear, Calle 30 No 502, Miramar, Ciudad Habana, C.P. 11300, Cuba}
\email{gran@ceaden.edu.cu}

\author{Augusto Gonz\'alez}
\affiliation{Instituto de Cibern\'etica, Matem\'atica y F\'{\i}sica, Calle
 E 309, Vedado, Ciudad Habana, Cuba}
\email{agonzale@icmf.inf.cu}

\author{D.J. Lockwood}
\affiliation{Institute for Microstructural Sciences, National Research Council, Ottawa, Canada K1A 0R6}
\email{David.Lockwood@nrc-cnrc.gc.ca}

\begin{abstract}
En este art\'iculo presentamos c\'alculos te\'oricos de la secci\'on eficaz de dispersi\'on inel\'astica de luz (Raman) por excitaciones electr\'onicas en un punto cu\'antico cargado con 42 electrones. Los estados multielectr\'onicos involucrados en el c\'alculo de la amplitud de transici\'on Raman son obtenidos en la aproximaci\'on de fase aleatoria (RPA). La evaluaci\'on de las reglas de sumas pesadas por energ\'ia permiti\'o clasificar las excitaciones electr\'onicas en uniparticulares (SPEs) y colectivas de carga (CDEs) y de esp\'in (SDEs). Los espectros Raman calculados en geometr\'ia polarizada y no polarizada son mostrados en diferentes regiones de la energ\'ia del laser incidente. Se predicen te\'oricamente las ventajas asociadas al c\'alculo o medici\'on de los espectros en r\'egimen no resonante. Los valores calculados de las razones de polarizaci\'on de las intensidades Raman demuestran el rompimiento de las reglas de selecci\'on del efecto Raman en presencia de campos magn\'eticos externos. La regla del salto de la intensidad Raman es propuesta como herramienta novedosa para la identificaci\'on de la naturaleza (carga o esp\'in) de las excitaciones electr\'onicas. En la regi\'on de resonancia extrema, se obtienen espectros Raman dominados por picos asociados a estados excitados uniparticulares. Se argumenta cualitativamente la importancia de incorporar los tiempos de vida de los estados intermedios para la descripci\'on de los espectros Raman en la regi\'on de energ\'ias del laser incidente $40\sim 50$ meV por encima de la brecha energ\'etica de la estructura semiconductora. 
\end{abstract}

\maketitle

\section{Introducci\'on}

En la actualidad un elevado porciento de las investigaciones de la f\'isica de la materia condensada, concentran sus esfuerzos en el estudio de las propiedades f\'isicas de heteroestructuras semiconductoras cuyas dimensiones espaciales se encuentran en la escala nanom\'etrica. T\'ecnicas modernas sofisticadas como la Epitaxia de Haces Moleculares (MBE) han hecho posible la obtenci\'on de gases electr\'onicos cuasi-bidimensionales en materiales semiconductores. Tales sistemas constituyen un punto de partida ideal para la fabricaci\'on de estructuras con dimensionalidad reducida como los hilos (cuasi-unidimensionales) y los puntos (cuasi-cerodimensionales) cu\'anticos  \cite{heitmann_phys_today} . Estos \'ultimos, conocidos tambien como \'atomos artificiales dada la cuantizaci\'on completa de su espectro energ\'etico, son obtenidos por diversos m\'etodos y con diversas geometr\'ias \cite{libro_hawrylak}. El estudio de las propiedades \'opticas y electr\'onicas de los puntos cu\'anticos ha tenido en los \'ultimos a\~nos una gran fuerza \cite{proccedings_physE_qdots2004} dado que los resultados de estas investigaciones tienen un impacto directo en el desarrollo de nuevas ramas de la f\'isica aplicada.

La espectroscop\'ia de excitaciones electr\'onicas en \'atomos artificiales, o sea, puntos cu\'anticos que contienen un n\'umero $N$ de electrones confinados en la banda de conducci\'on, constituye la analog\'ia m\'as natural de la espectroscop\'ia en sistemas at\'omicos. El espectro de estados excitados de \'atomos artificiales es una informaci\'on b\'asica para el desarrollo de dispositivos electr\'onicos cuyos principios de funcionamiento se basan en los efectos cu\'anticos presentes en estas nanoestructuras. La dispersi\'on inelastica de luz (Raman) permite investigar diferentes sectores del espectro de excitaciones electr\'onicas de estos sistemas. En la d\'ecada de los a\~nos 1990 aparecieron en la literatura los primeros trabajos experimentales sobre mediciones de espectros Raman asociados a excitaciones electr\'onicas en \'atomos artificiales que contienen cientos de electrones \cite{strenz_prl_1994,shuller_PRB_1996,lockwood_PRL_1996,shuller_PRL_1998}. Una amplia descripci\'on de los resultados obtenidos en estos trabajos aparece publicada en las referencias \cite{Lockwood_review_2000, shuller_ssc_2001}. Recientemente se han reportado mediciones de espectros Raman de excitaciones electr\'onicas en puntos cu\'anticos notablemente peque\~nos que contienen como promedio entre 2 y 6 electrones confinados en las dimensiones de la estructura semiconductora \cite{heitmann_PRL_2003, Molinari_cond_matt_2005}.

En general las excitaciones electr\'onicas en puntos cu\'anticos pueden clasificarse en excitaciones uniparticulares (SPEs) y excitaciones colectivas de carga (CDEs) y de esp\'in (SDEs).  Los picos Raman asociados a las CDEs pueden ser medidos o calculados en la llamada configuracion de geometr\'ia polarizada, en la cual los vectores de campo el\'ectrico del fot\'on dispersado e incidente son paralelos mientras que los estados excitados del tipo SDEs son obtenidos en geometr\'ia no polarizada donde el vector de campo el\'ectrico del fot\'on dispersado es ortogonal con respecto a la misma magnitud del fot\'on incidente. Esta diferenciaci\'on esta dada por las reglas de selecci\'on del efecto Raman \cite{ORA_PHYE_2004}. Sch\"uller y colaboradores \cite{shuller_PRB_1996, shuller_PRL_1998} han mostrado experimentalmente que fuertes picos Raman asociados a SPEs aparecen en los espectros medidos como resultado de iluminar el punto con un laser cuya energ\'ia esta muy cercana al valor de la brecha energ\'etica ($E_{\rm{gap}}$) del semiconductor mientras que para energ\'ias de incidencias muy por encima de $E_{\rm{gap}}$ el espectro Raman es dominado por excitaciones de naturaleza colectiva.

La obtenci\'on de estos resultados experimentales motiv\'o un conjunto de estudios te\'oricos  encaminados a la despcripci\'on de la fenomenolog\'ia encontrada en los citados trabajos. Estas teor\'ias, \cite{reboredo,Brataas,steinebach,barranco,lipparini} s\'olo v\'alidas en r\'egimen no resonante, han descrito las posiciones de las excitaciones colectivas para diferentes valores del vector de onda de la radiaci\'on incidente y de campos magn\'eticos externos. Sin embargo, los citados formalismos presentan limitaciones inherentes a las aproximaciones usadas tales como: i) la incapacidad de describir las relaciones de intensidades entre los picos Raman asociados a SPEs y CDEs y/o SDEs obtenidas en los experimentos, ii) la incapacidad de predecir la presencia de picos Raman asociados a SPEs, iii) no consideran los estados electr\'onicos en la banda de valencia, los cuales juegan un papel crucial en el c\'alculo de la secci\'on eficaz de dispersi\'on Raman \cite{AAL_PRB_2004}. Este art\'iculo tienen como objetivo fundamental enriquecer y complementar las interpretaciones discutidas en los trabajos experimentales publicados en esta tem\'atica aunque no s\'olo se limita a eso dado que nuevas predicciones te\'oricas han emergido del an\'alisis de los c\'alculos realizados.

En la pr\'oxima secci\'on describimos de manera general el esquema de c\'alculo implementado para la obtenci\'on de la secci\'on eficaz diferencial Raman. La secci\'on \ref{results} contiene la discusi\'on y an\'alisis de los resultados. Finalmente las conclusiones generales son brindadas en la secci\'on \ref{conclusions}.

\section{M\'etodos Te\'oricos y esquema de c\'alculo implementado}

\begin{figure}[ht]
\begin{center}
\includegraphics[width=1\linewidth,angle=0]{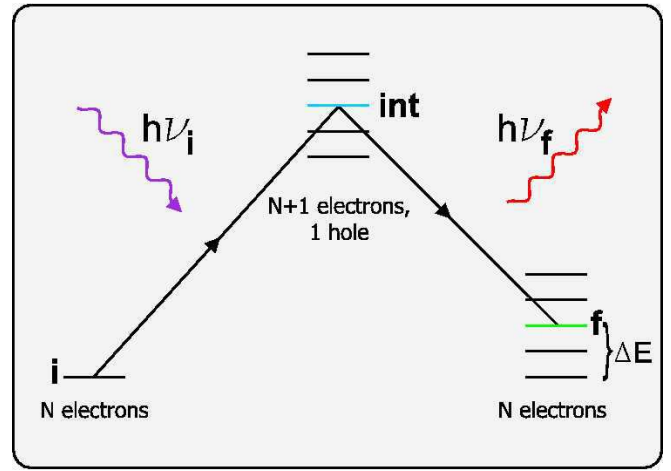}
\caption{\label{fig2_ssc_aal2005} Amplitud de transici\'on Raman y su interpretaci\'on en t\'erminos de transiciones virtuales.}
\end{center}
\end{figure}

La idea b\'asica de la dispersi\'on Raman es simple: un fot\'on con energ\'ia $h\nu_i$ el cual es dispersado inel\'asticamente por la materia puede ganar (procesos anti-Stoke) o perder (Stoke) energ\'ia. La teor\'ia de perturbaciones permite obtener la expresi\'on para el c\'alculo de la amplitud de transici\'on Raman \cite{Loudon_text} $A_{fi}$, definida por la siguiente ecuaci\'on:

\begin{equation}
A_{fi}\sim \sum_{int} \frac{\langle f|\hat H^+_{e-r}|int\rangle
\langle int|\hat H^-_{e-r}|i\rangle}{h\nu_i-(E_{int}-E_i)+i\Gamma_{int}}.
\label{eq1}
\end{equation} 
Una representaci\'on esquem\'atica del proceso de dispersi\'on Raman es mostrado en la figura \ref{fig2_ssc_aal2005}.
$|i\rangle$ y $|f\rangle$ son los estados electr\'onicos inicial y final respectivamente que describen a los $N$-electrones confinados en el punto cu\'antico. En principio, la suma en la Ec. (\ref{eq1}) debe incluir todos los estados electr\'onicos que no anulen los elementos matriciales en el numerador. En los trabajos experimentales mencionados la energ\'ia del laser incidente, $h\nu_{i}$, toma valores del orden de $E_{\rm{gap}}$. Esto permite restringir la suma en la Ec. (\ref{eq1}) por estados intermedios que son excitaciones interbandas de la estructura semiconductora, o sea, por estados de un sistema en el cual existen $(N+1)$ - electrones confinados en la banda de conducci\'on y $1$ h confinado en la banda de valencia. $\hat H_{e-r}$ denota el Hamiltoniano de interacci\'on del sistema electr\'onico con el campo de fotones de la radiaci\'on incidente \cite{ORA_PHYE_2004}. $E_i$ y $E_{\rm{int}}$ son las energ\'ias correspondientes a los estados inicial e intermedios y $\Gamma_{\rm{int}}$ es un par\'ametro asociado a los tiempos de vida de estos \'ultimos. La ley de conservaci\'on de la energ\'ia establece la ecuaci\'on,

\begin{equation}
h\nu_i-h\nu_f=E_f-E_i=\Delta E,
\label{energy}
\end{equation} 
donde $h\nu_f$ es la energ\'ia del fot\'on dispersado, $E_f$ es la energ\'ia del estado excitado final y $\Delta E$ es la energ\'ia de excitaci\'on electr\'onica o corrimiento Raman asociado al estado $|f\rangle$. Una vez obtenida la amplitud de transici\'on Raman podemos calcular la secci\'on eficaz diferencial de dispersi\'on Raman, $d\sigma$, a trav\'es de la siguiente expresi\'on:

\begin{equation}
d\sigma\sim\sum_f |A_{fi}|^2 \delta(E_i+h\nu_i-E_f-h\nu_f).
\label{cross_section}
\end{equation}

La conservaci\'on de la energ\'ia es forzada por la funci\'on delta de Dirac en la Ec. \ref{cross_section}. En nuestros c\'alculos, \'esta es simulada por la funci\'on de Lorentz

\begin{equation}
\delta(x)\approx \frac{\Gamma_f/\pi}{x^2+\Gamma_f^2}.
\end{equation}
\begin{figure*}[ht]
\begin{center}
\includegraphics[width=1\linewidth,angle=0]{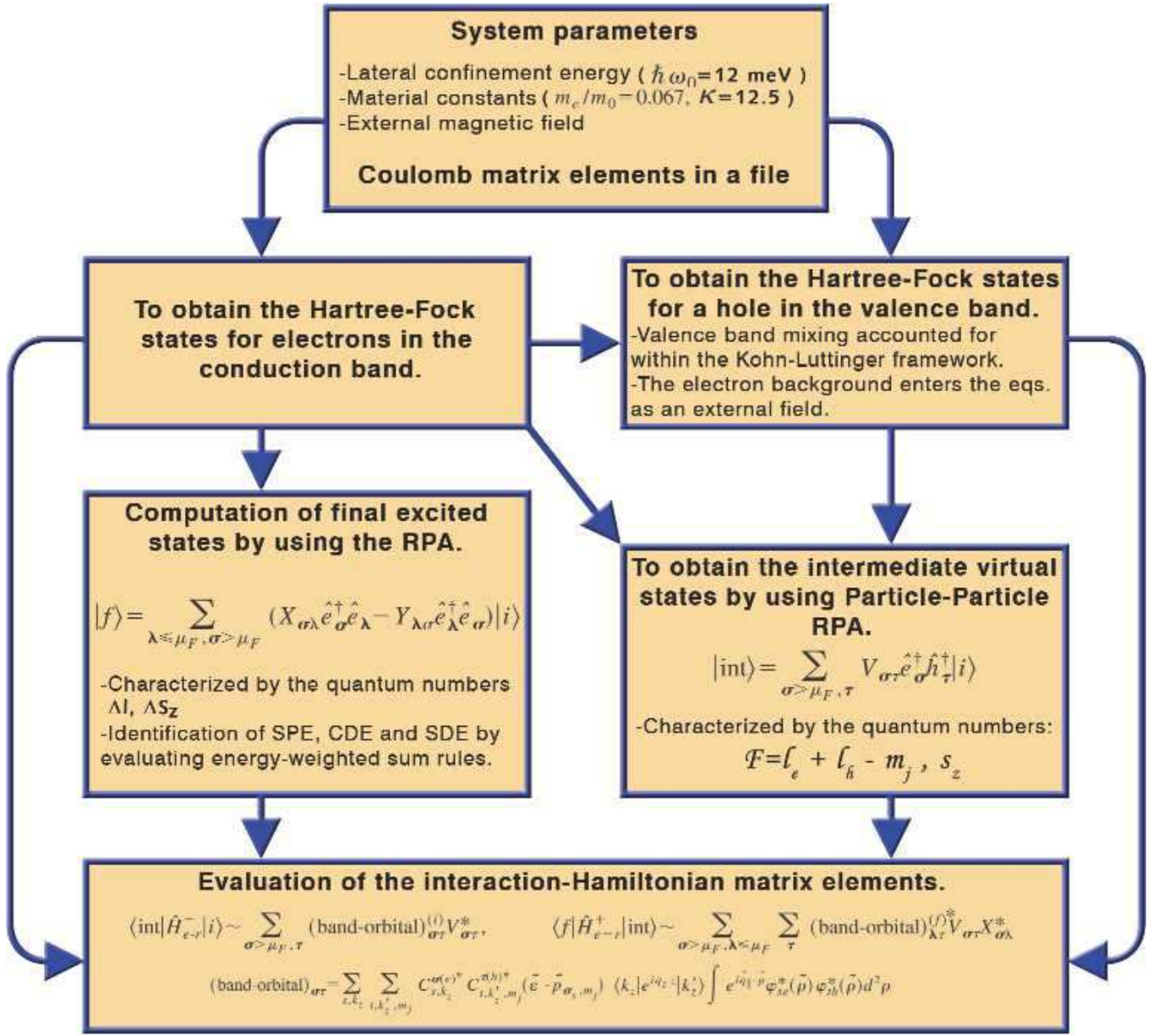}
\caption{\label{fig3_ssc_aal2005} Esquema de c\'alculo implementado}
\end{center}
\end{figure*}

\subsection{Esquema computacional para la obtenci\'on de la amplitud de transici\'on Raman}

Las ecuaciones y expresiones expl\'icitas detalladas de las diferentes aproximaciones involucradas en el c\'alculo de la amplitud de transici\'on Raman, $A_{fi}$, pueden ser encontradas por los lectores interesados en las referencias [\onlinecite{AAL_PRB_2004, AAE_PRB_2002}]. En esta secci\'on brindaremos una descripci\'on de la metodolog\'ia de c\'alculo implementada auxili\'andonos de la figura \ref{fig3_ssc_aal2005}.

El punto cu\'antico en forma de disco es modelado con un potencial de paredes r\'igidas en la direcci\'on $z$ (direcci\'on de crecimiento de la heteroestructura) y un confinamiento lateral (direcciones $x, y$) de tipo parab\'olico con una energ\'ia caracter\'istica $\hbar\omega_0 = 12$ meV. La base de funciones utilizada que caracteriza los estados de una part\'icula est\'an constru\'idas como el producto de las autofunciones del oscilador bidimensional, del pozo infinito y las funciones de esp\'in \cite{AAL_PRB_2004}. 

Con el objetivo de hacer eficiente el c\'alculo de las energ\'ias y funciones de onda de los sistemas multielectr\'onicos involucrados, nosotros calculamos previamente todos los valores de los elementos matriciales de la interacci\'on de Coulomb $\langle\alpha,\beta|1/r|\gamma,\delta\rangle$ donde $\alpha$, $\beta$, $\gamma$, $\delta$, son estados arbitrarios de una part\'icula. Estos elementos matriciales son cargados en memoria al inicio de cada c\'alculo permiti\'endonos resolver las ecuaciones no lineales integro-diferenciales de Hartree-Fock para 42-electrones en pocos minutos, o calcular todos los estados intermedios considerados (alrededor de 10,000) en la Ec. (\ref{eq1}) en pocos d\'ias. Las ecuaciones de Hartree-Fock para huecos incluye el campo electrost\'atico generado por el {\it background} de electrones en el punto y considera los efectos de la mezcla de las bandas de valencia a trav\'es del Hamiltoniano de Kohn-Luttinger \cite{AAL_PRB_2004}.

Los estados excitados finales e intermedios son descritos en el marco de la Aproximacion de Fase Aleatoria \cite{Ring} (RPA). Mediante la RPA es posible calcular las energ\'ias y funciones de onda de los estados finales en diferentes sectores del espectro de excitaciones electr\'onicas del punto, donde cada sector est\'a definido por reglas de selecci\'on para la variaci\'on (respecto al valor de la magnitud en el estado base) del momentum angular ($\Delta l=0, 1, 2,..$) y la proyecci\'on del esp\'in ($\Delta S_z=0,\pm 1$). La evaluaci\'on de reglas de sumas pesadas por energ\'ia \cite{Ring} permite identificar de manera independiente al c\'alculo de la amplitud de transici\'on Raman los estados colectivos de carga (CDEs), de esp\'in (SDEs) y las excitaciones uniparticulares (SPEs). Por otro lado, los estados intermedios del proceso Raman, est\'an caracterizados por la proyecci\'on del esp\'in del electr\'on a\~nadido $S_z$ y el momento angular total del par, ${\cal F}=l_e+l_h-m_j$.

Una vez calculados los estados finales e intermedios, son evaluados los elementos matriciales del Hamiltoniano de interacci\'on del punto cu\'antico con el campo electromagn\'etico en t\'erminos de los coeficientes que caracterizan las funciones de onda de los estados multielectr\'onicos finales e intermedios y los coeficientes de la expansi\'on de los estados uniparticulares de Hartree-Fock de electrones y huecos en la bases de funciones utilizada.

\subsection{Aproximaci\'on para el c\'alculo de la amplitud de transici\'on Raman en r\'egimen no resonante (ORA)}

La ORA ({\it off-resonant approximation}) es una aproximaci\'on en la cual la expresi\'on para el c\'alculo de la amplitud de transici\'on Raman, $A_{fi}$, involucra solamente a los estados final e inicial del punto cu\'antico. La suma por estados intermedios desaparece; esto significa que en los marcos de esta aproximaci\'on no es posible la explicaci\'on de efectos producto de resonancias con estados intermedios en el proceso de dispersi\'on Raman.

La deducci\'on de la expresi\'on para la amplitud de transici\'on Raman en esta aproximaci\'on puede ser consultada en la referencia [\onlinecite{ORA_PHYE_2004}]. Las suposiciones b\'asicas para su derivaci\'on son: (i) la energ\'ia del l\'aser incidente est\'a lo suficientemente lejos de las energ\'ias de los estados intermedios, por lo que la dependencia en la Ec. (\ref{eq1}) con $E_{\rm{int}}$ en el denominador puede ser despreciada; (ii) existe una ventana energ\'etica de estados intermedios (40 meV por encima de $E_{\rm{gap}}$) donde las variaciones de $\Gamma_{\rm{int}}$ son despreciables y la relaci\'on de completitud $\sum_{int}' |int\rangle\langle int|\approx 1$ es casi satisfecha. Bajo estas suposiciones la amplitud de transici\'on Raman, $A_{\rm{fi}}^{\rm{ORA}}$, puede escribirse como:

\begin{equation}
A_{fi}^{ORA}\sim\langle f|H^+_{e-r}H^-_{e-r}|i\rangle.
\label{eq_ora1}
\end{equation}

Haciendo uso de las expresiones para $H^+_{e-r}$ y $H^-_{e-r}$ \cite{ORA_PHYE_2004} es posible deducir la siguiente formula para el c\'alculo de de la amplitud de transic\'on Raman:
\begin{widetext}
\begin{eqnarray}
A_{fi}^{ORA}&\sim&\sum_{\alpha,\alpha'}
 \langle\alpha| e^{i(\vec q_i-\vec q_f)\cdot \vec r}|\alpha'\rangle
 \left\{\frac{2}{3} (\vec \varepsilon_i\cdot\vec\varepsilon_f) 
 \left\langle f\left|~
 e^\dagger_{\alpha\uparrow} e_{\alpha'\uparrow}+
 e^\dagger_{\alpha\downarrow} e_{\alpha'\downarrow} 
 \right.\right|i\right\rangle\nonumber\\
&+&\frac{i}{3} (\vec \varepsilon_i\times\vec\varepsilon_f)\cdot 
 \left.\left\langle f\left|~
 \hat z~(e^\dagger_{\alpha\uparrow} e_{\alpha'\uparrow}-
 e^\dagger_{\alpha\downarrow} e_{\alpha'\downarrow})+
 (\hat x+i\hat y)~  e^\dagger_{\alpha\uparrow} e_{\alpha'\downarrow}+
 (\hat x-i\hat y)~ e^\dagger_{\alpha\downarrow} e_{\alpha'\uparrow}
    \right|i\right\rangle\right\}.
\label{eq6}
\end{eqnarray}
\end{widetext}

Del an\'alisis de la Ec. (\ref{eq6}) podemos inferir algunas conclusiones. Primero, solamente los estados finales excitados colectivos tendr\'an una amplitud Raman asociada diferente de zero. No es posible describir picos Raman asociados a excitaciones uniparticulares en esta aproximaci\'on. Segundo, las reglas de selecci\'on del efecto Raman asociadas a las polarizaciones de los fotones incidente y dispersado puede ser entendida de la Ec. (\ref{eq6}): el primer t\'ermino, el cual no altera el n\'umero cu\'antico de esp\'in del estado inicial est\'a multiplicado por un factor $\vec \varepsilon_i\cdot\vec\varepsilon_f$. Esto significa que los estados finales colectivos de carga aparecen en geometr\'ia polarizada. Por otro lado, los otros t\'erminos los cuales modifican el esp\'in del estado inicial est\'an multiplicados por un factor $\vec \varepsilon_i\times\vec\varepsilon_f$ y consecuentemente los picos Raman asociados a estas excitaciones aparecen en geometr\'ia no polarizada.

\section{Resultados}
\label{results}
En este art\'iculo reportamos los resultados para un punto cu\'antico con $42$ electrones confinados en la banda de conducci\'on. El ancho del pozo en la direcci\'on $z$ es $L=25$ nm y la energ\'ia de confinamiento caracter\'istica es $\hbar\omega_0=12$ meV.
\begin{figure*}[ht]
\begin{center}
\includegraphics[width=.5\linewidth,angle=-90]{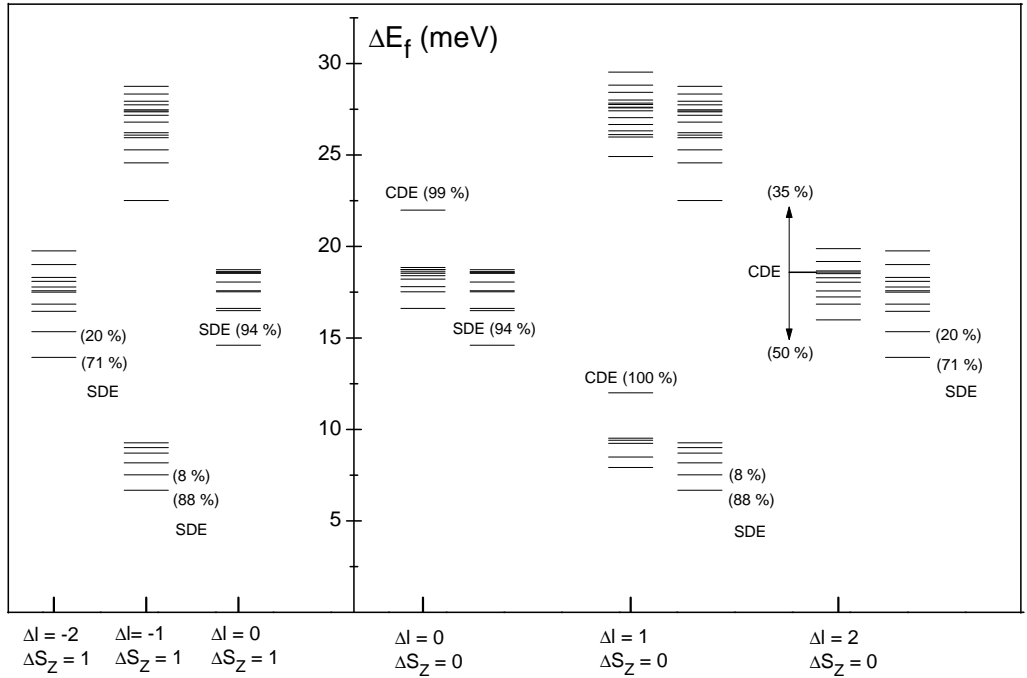}
\caption{\label{fig2_prb_aal2004} Espectro de excitaciones electr\'onicas multipolares del punto cu\'antico. $B=0$ T.}
\end{center}
\end{figure*}
El n\'umero de electrones corresponde a un punto de capas cerradas. En nuestro estudio consideramos estados excitados con energ\'ias de excitaci\'on $\Delta E \leq\hbar\omega_{\rm{LO}}$, siendo $\hbar\omega_{\rm{LO}}\approx 30$ meV el valor de energ\'ia umbral para la emisi\'on de fonones \'opticos longitudinales en el GaAs. Esto significa que los estados finales tienen peque\~nos anchos para los cuales hemos tomado un valor constante $\Gamma_f=0.1$ meV. De igual manera para los estados intermedios con energ\'ias de excitaci\'on menores que $\hbar\omega_{\rm{LO}}$ escogemos el par\'ametro $\Gamma_{\rm{int}}=$0.5 meV. Para valores superiores en energ\'ias de excitaci\'on, como resultado de la emisi\'on de fonones LO fijamos el par\'ametro $\Gamma_{\rm{int}}=10$ meV. Todos los espectros Raman est\'an calculados en la configuraci\'on de retrodispersi\'on donde la luz incidente y dispersada forman un \'angulo de $20^{\rm{o}}$ respecto a la normal del punto.

\subsection{Espectro de excitaciones electr\'onicas}
En la figura \ref{fig2_prb_aal2004} mostramos el espectro de estados finales excitados del punto modelado calculados en la aproximaci\'on de fase aleatoria. Las energ\'ias de excitaci\'on de estos estados es precisamente la magnitud medida en los experimentos de dispersi\'on inel\'astica de luz (corrimiento Raman). En la figura se identifican los estados colectivos de carga (CDE) y esp\'in (SDE) en todos los sectores y la contribuci\'on de cada estado colectivo a la regla de suma pesada por energ\'ia es expl\'icitamente se\~nalada. El resto de los estados mostrados corresponden a excitaciones uniparticulares. 

\begin{figure}[ht]
\begin{center}
\includegraphics[width=1\linewidth,angle=-90]{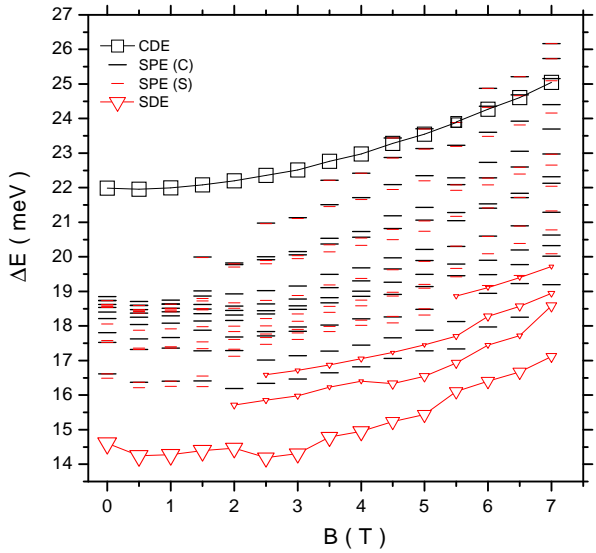}
\caption{\label{fig1_prbrc_aal2005} Espectro de excitaciones electr\'onicas monopolares ($\Delta l=0, \Delta S_z =0$) vs campo magn\'etico externo.}
\end{center}
\end{figure}

La evoluci\'on del espectro de estados excitados monopolares ($\Delta l=0, \Delta S_z =0$) como funci\'on del campo magn\'etico externo es mostrada en la figura \ref{fig1_prbrc_aal2005}. Las excitaciones colectivas de carga y esp\'in cuya contribuci\'on a la regla de suma pesada por energ\'ia excede el 5 \% son representados con cuadrados y tri\'angulos respectivamente. La contribuci\'on de cada modo es proporcional al tama\~no de los s\'imbolos usados. Las excitaciones uniparticulares de carga (SPEs(C)) y esp\'in (SPEs(S)) son tambi\'en mostradas con l\'ineas horizontales largas y cortas respectivamente. 

\begin{figure*}[ht]
\begin{center}
\includegraphics[width=.7\linewidth,angle=-90]{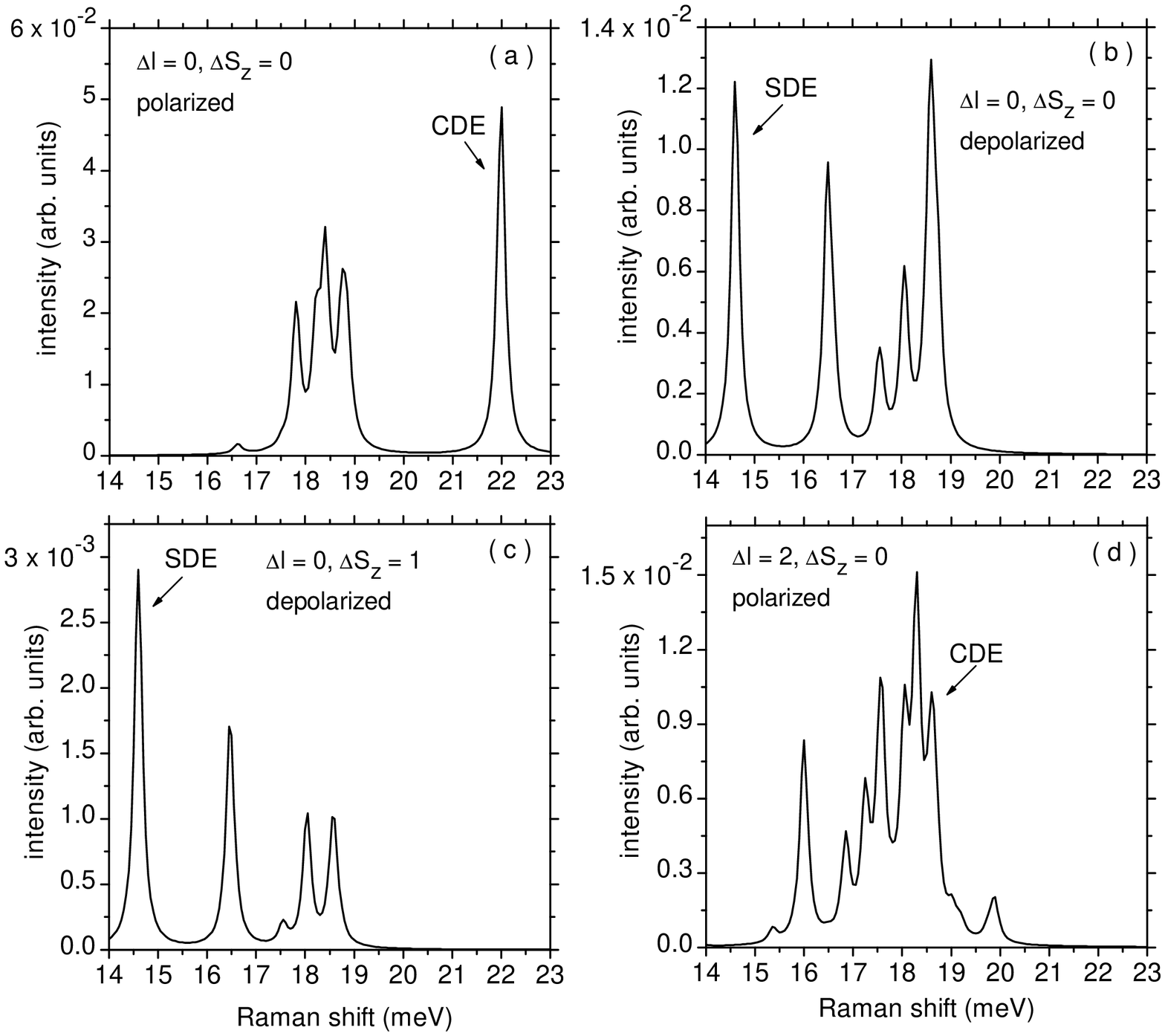}
\caption{\label{fig4_ssc_aal2005} Espectros Raman para diferentes sectores de excitaciones electr\'onicas. $h\nu_i = E_{\rm{gap}}-5$ meV, $B=0$ T.}
\end{center}
\end{figure*}
Con relaci\'on a las excitaciones colectivas de carga se pueden notar dos caracter\'isticas fundamentales: (i) la energ\'ia de excitaci\'on asociada a este modo evoluciona de manera suave respecto al campo magn\'etico externo y (ii) el tama\~no de los s\'imbolos no cambia para diferentes valores del campo, manifest\'andose la fuerte naturaleza colectiva de estos modos.

Los estados de bajas energ\'ias de excitaci\'on corresponden a excitaciones colectivas monopolares de esp\'in. A diferencia de los modos colectivos de carga, podemos observar para valores de campo magn\'eticos $B \geq 2$ T m\'as de un estado colectivo contribuyendo a la regla de suma pesada por energ\'ia.  

\subsection{Espectros Raman en r\'egimen no resonante ($h\nu_i<E_{\rm{gap}}$)} 

En este r\'egimen no existen resonancias con estados intermedios. Sin embargo, esto no implica que la f\'isica de la espectroscop\'ia Raman pueda ser descrita en su totalidad con una aproximaci\'on que desprecie la contribuci\'on de dichos estados como la ORA. Las principales caracter\'isticas de los espectros Raman en esta regi\'on son expuestos de manera resumida a continuaci\'on. Un an\'alisis mas amplio de los mismos puede ser consultado en la referencia [\onlinecite{ssc_aal_2005}].

\subsubsection{Contribuci\'on de las excitaciones electr\'onicas multipolares en los espectros Raman}
Espectros Raman asociados a excitaciones electr\'onicas en diferentes sectores de momentun angular y esp\'in son mostrados en la figura \ref{fig4_ssc_aal2005}. Las intensidades Raman correspondientes a excitaciones monopolares y cuadrupolares ($\Delta l = 2$) exhiben magnitudes comparables aunque se observa un predominio de los picos Raman asociados a los modos monopolares.
La contribuci\'on al espectro Raman en geometr\'ia no polarizada de las excitaciones de esp\'in del tipo {\it spin-flip} ($\Delta S_z=\pm 1$) es mucho mas d\'ebil comparada con la contribuci\'on de los estados excitados de esp\'in caracterizados por un cambio en el valor del esp\'in total pero no en el valor de su proyecci\'on. Dado que ambos modos est\'an muy cercanos en energ\'ias los picos Raman colectivos medidos en geometr\'ia no polarizada son adjudicables a estados excitados en los cuales la proyecci\'on total del esp\'in del estado base no es modificada en el proceso de dispersi\'on. Los espectros Raman asociados a excitaciones dipolares ($\Delta l = 1)$ \cite{AAL_PRB_2004} revelan intensidades uno o dos \'ordenes de magnitud menores en comparaci\'on con las mostradas en la figura \ref{fig4_ssc_aal2005}.  

\subsubsection{Comportamiento de las intensidades Raman como funci\'on de la energ\'ia del l\'aser incidente}
\label{smooth_bvelow_gap}

\begin{figure}[ht]
\begin{center}
\includegraphics[width=1\linewidth,angle=-90]{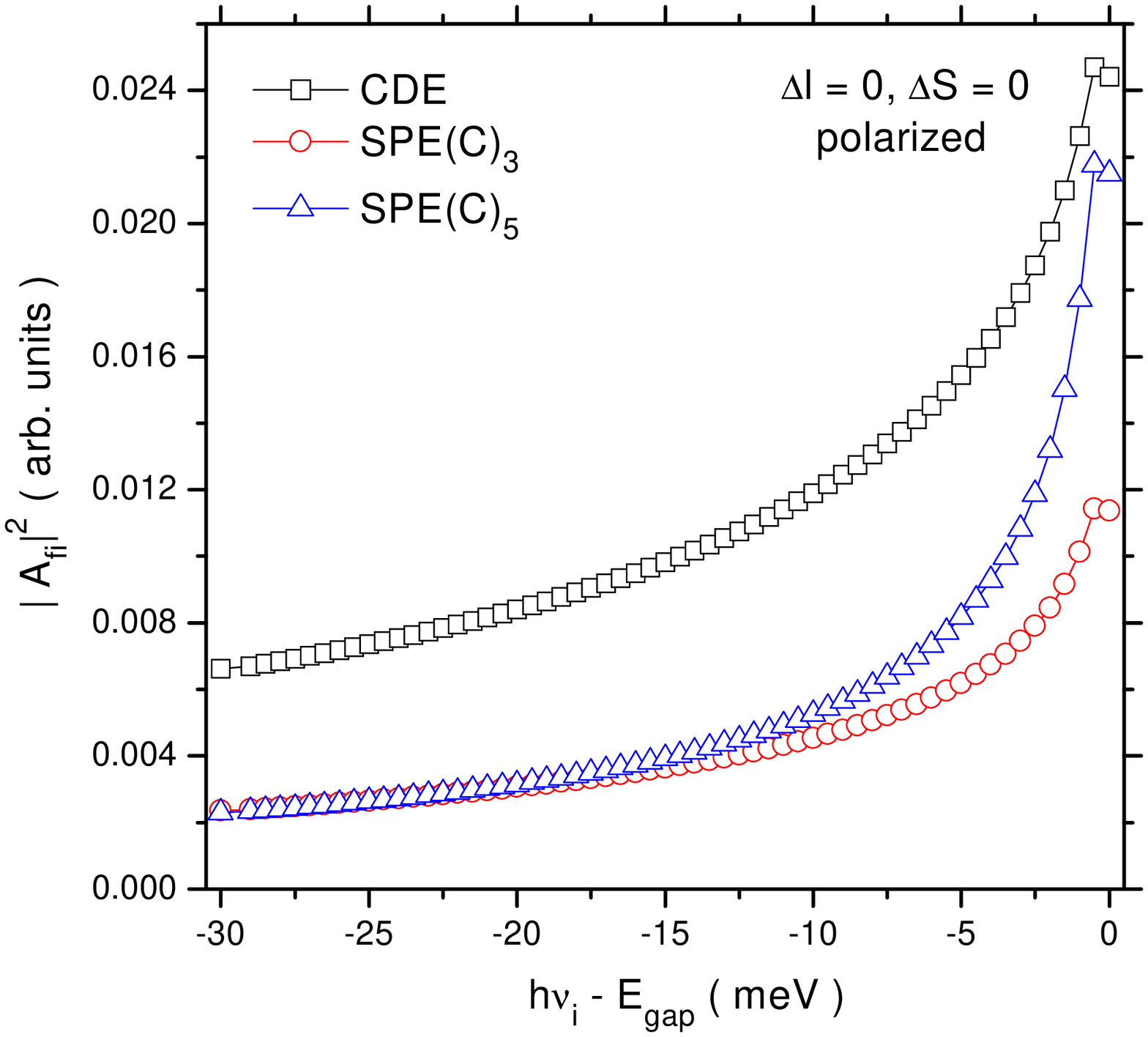}
\caption{\label{fig5_ssc_aal2005} Comportamiento de $\vert A_{\rm{fi}}\vert ^2$ para el modo colectivo monopolar y dos modos uniparticulares de carga como funci\'on de la energ\'ia del laser incidente. $B=0$ T.}
\end{center}
\end{figure}

Es posible a partir del esquema de c\'alculo implementado monitorear el valor de la intensidad Raman asociada a cualquier estado excitado del punto cu\'antico, en la medida que se var\'ia la energ\'ia del l\'aser incidente. En esta regi\'on, donde la energ\'ia de la luz que incide sobre el punto es variada en el intervalo ($E_{\rm{gap}}-30$ meV, $E_{\rm{gap}}$), las intensidades Raman asociadas tanto a los modos SPEs como a los modos colectivos (CDEs, SDEs) muestran un comportamiento mon\'otono creciente cuando $h\nu_i$ se incrementa. Este comportamiento se muestra en la figura \ref{fig5_ssc_aal2005}
donde hemos ploteado los valores de $\vert A_{\rm{fi}}\vert ^2$ para dos estados uniparticulares y el modo colectivo del sector monopolar de excitaciones electr\'onicas. Aunque no se han reportado mediciones de espectros Raman en esta regi\'on, desde el punto de vista te\'orico se predice un comportamiento de las intensidades Raman muy conveniente para la identificaci\'on de los picos en un experimento real. Otro aspecto a resaltar es que, efectivamente, existe un l\'imite para el valor de energ\'ia del laser incidente por debajo de la brecha energ\'etica del semiconductor a partir del cual los picos mas fuertes en el espectro est\'an asociados a excitaciones colectivas. Dicho l\'imite, a partir del cual la fenomenolog\'ia del efecto Raman puede ser descrita con una aproximaci\'on del tipo de la ORA fue investigado en la referencia [\onlinecite{ORA_PHYE_2004}].

\begin{figure}[ht]
\begin{center}
\includegraphics[width=1\linewidth,angle=0]{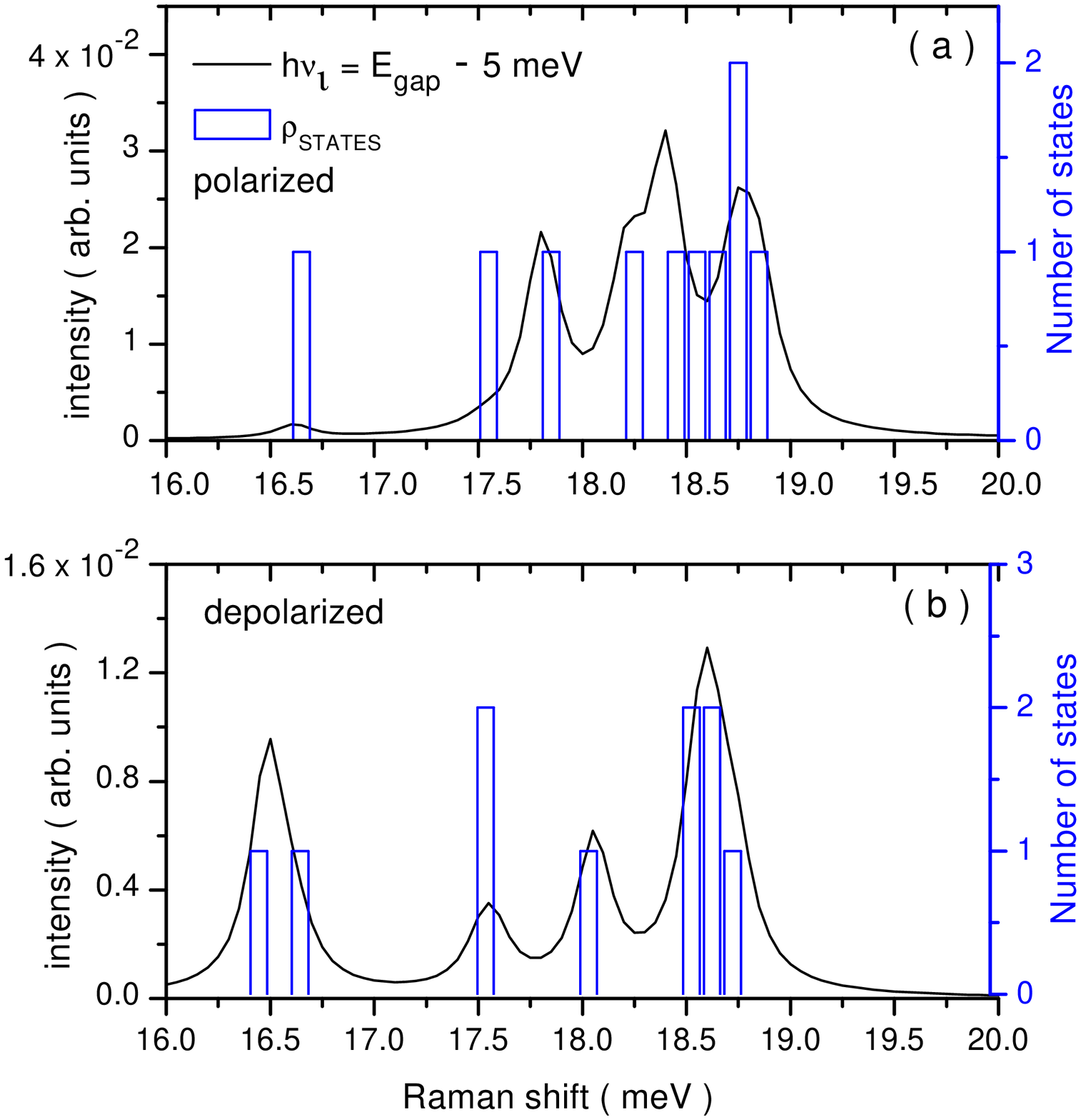}
\caption{\label{fig6_ssc_aal2005} Espectros Raman de excitaciones monopolares en ambas geometr\'ias. Comparaci\'on con la densidad de estados uniparticulares finales. $B=0$ T.}
\end{center}
\end{figure}

\subsubsection{Correlaci\'on entre las intensidades de los picos Raman y la densidad de estados finales}
Es razonable esperar picos Raman intensos para valores de la energ\'ia de excitaci\'on donde existe una aglomeraci\'on de estados excitados finales de una part\'icula. Lo novedoso en este punto, mostrado para el caso de excitaciones monopolares en la figura \ref{fig6_ssc_aal2005}, es la estrecha correlaci\'on obtenida entre la densidad de niveles de energ\'ia de los SPEs de carga con la intensidad de los picos Raman obtenidos en la configuraci\'on de geometr\'ia polarizada y entre los SPEs de esp\'in con la intensidad de los picos Raman obtenidos en la configuraci\'on de geometr\'ia no polarizada en ausencia de campos magn\'eticos externos. O sea, la correlaci\'on existente entre la intensidad de los picos Raman y la densidad de estados excitados uniparticulares est\'a mediada tambi\'en, en ausencia de campos magn\'eticos externos, por las reglas de selecci\'on del efecto Raman deducibles s\'olo para los modos colectivos.

\subsubsection{Rompimiento de las reglas de selecci\'on del efecto Raman en un campo magn\'etico externo}
\begin{figure}[ht]
\begin{center}
\includegraphics[width=1\linewidth,angle=0]{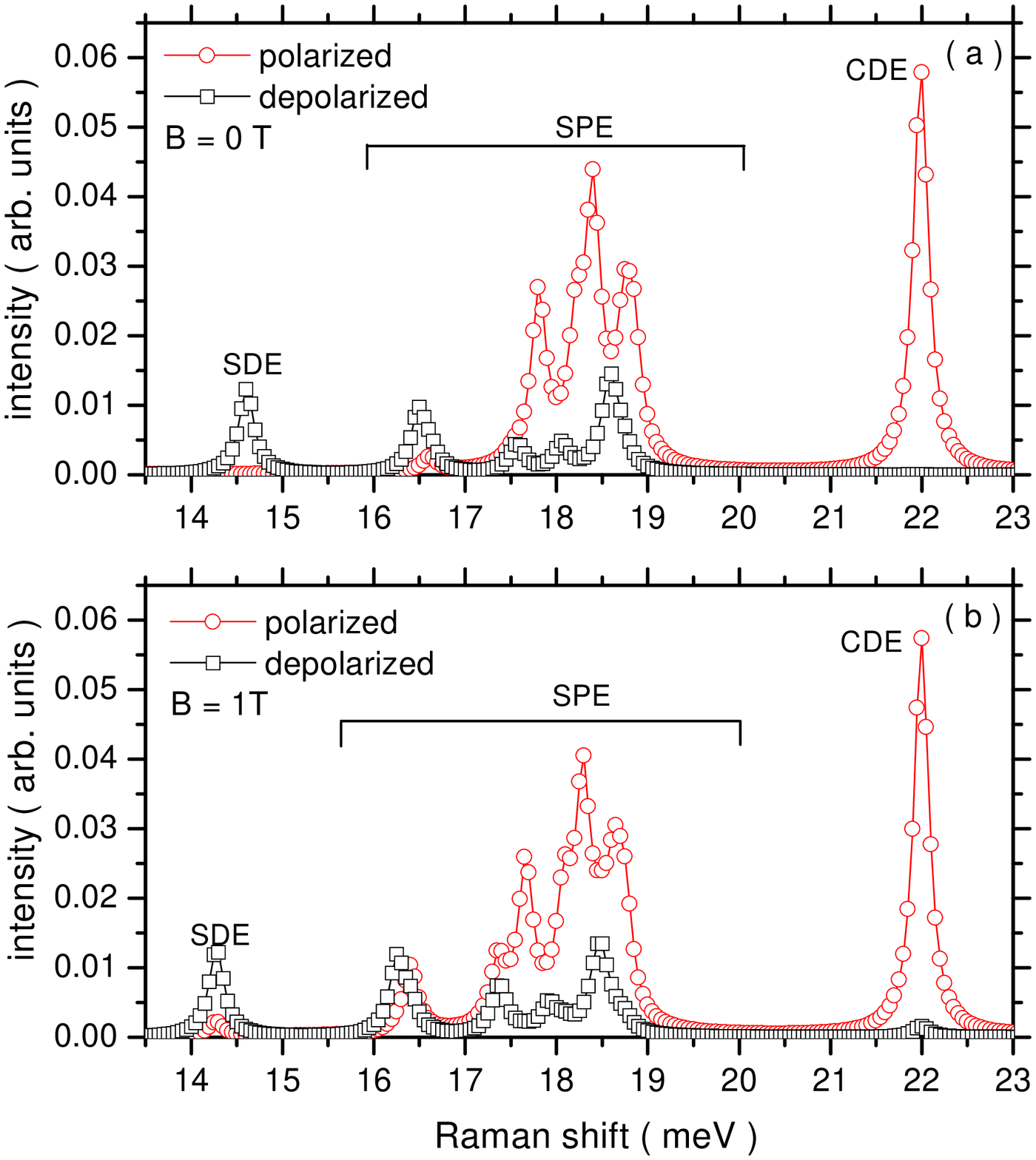}
\caption{\label{fig7_ssc_aal2005} Espectros Raman de excitaciones monopolares para $B=$0 y 1 T en ambas geometr\'ias.}
\end{center}
\end{figure}

En presencia de un campo magn\'etico externo perpendicular al plano de movimiento de los electrones confinados en el punto, las reglas de selecci\'on del efecto Raman se rompen parcialmente para los estados colectivos y dr\'asticamente para las excitaciones de una part\'icula. Este efecto es ilustrado en la figura \ref{fig7_ssc_aal2005} donde hemos ploteado los espectros Raman asociados a excitaciones monopolares para $B=0$ y $1$ T. Mediante el c\'alculo de las razones de polarizaci\'on \cite{AAL_PRBRC_2005} de los picos Raman monopolares para $B=0$, $1$ T se evalu\'o cuantitativamente los efectos del campo magn\'etico en las reglas de selecci\'on de la dispersi\'on Raman: (i) se chequearon para $B=0$ las reglas de selecci\'on para las intensidades Raman asociadas a los estados colectivos, (ii) el an\'alisis de las razones de polarizaci\'on de las intensidades Raman para cada modo uniparticular mostr\'o que dichas excitaciones, al igual que los modos colectivos, obedecen las reglas de selecci\'on del efecto Raman en ausencia de campos magn\'eticos externos, (iii) la presencia de un campo magn\'etico externo, en este caso $B=1$ T, rompe las reglas de selecci\'on del efecto Raman parcialmente para los modos colectivos y dr\'asticamente para los estados excitados uniparticulares. 

\subsubsection{Regla del salto de la intensidad Raman cuando $h\nu_i=E_{\rm{gap}}$}
\begin{figure}[ht]
\begin{center}
\includegraphics[width=1\linewidth,angle=0]{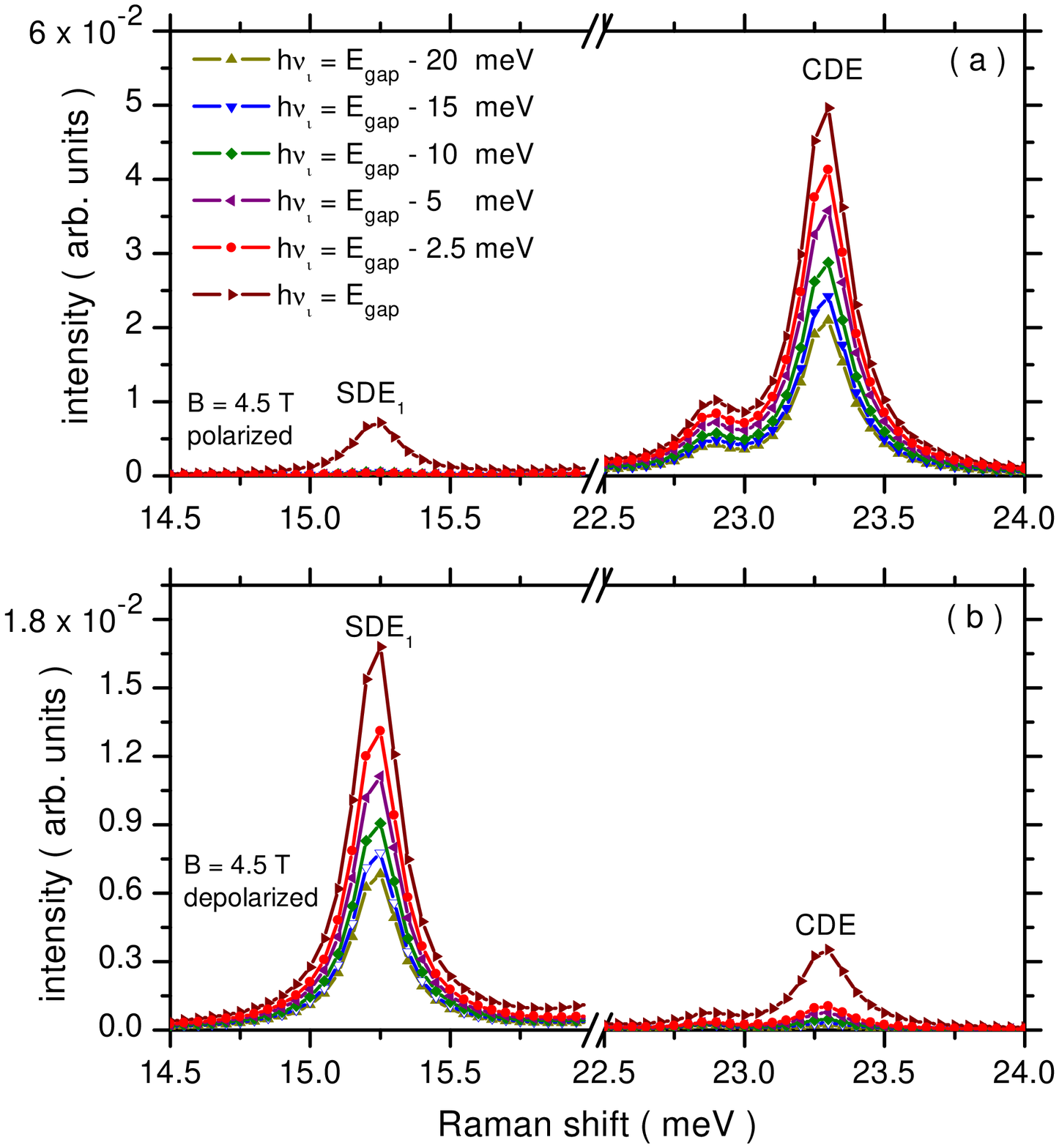}
\caption{\label{fig8_ssc_aal2005} Espectros Raman de excitaciones monopolares en ambas geometr\'ias para $B=$4.5 T. Los picos Raman asociados a las excitaciones uniparticulares han sido substra\'idos de los gr\'aficos.}
\end{center}
\end{figure}

\begin{figure}[ht]
\begin{center}
\includegraphics[width=1\linewidth,angle=0]{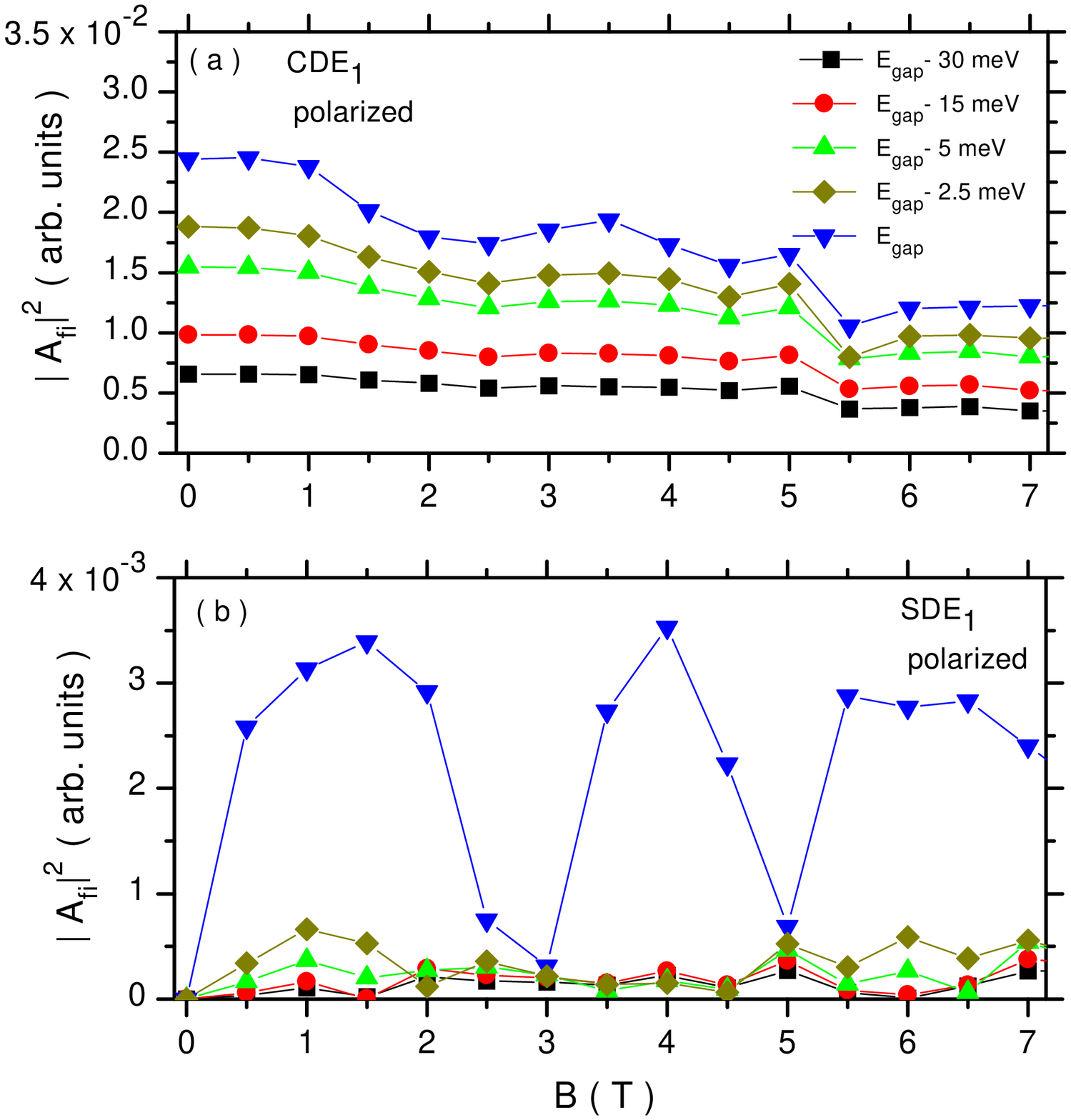}
\caption{\label{fig4_prbrc_aal2005} Comportamiento de las amplitudes de transici\'on Raman correspondientes a CDE (a) y $SDE_1$ (b) en geometr\'ia polarizada vs $B$.}
\end{center}
\end{figure}

Este es un resultado bien interesante de esta investigaci\'on  y est\'a estrechamente ligado con lo explicado en la secci\'on anterior. En presencia de un campo magn\'etico externo si seguimos el comportamiento de la intensidad Raman correspondiente a un estado excitado de carga (esp\'in) en geometr\'ia no polarizada (polarizada) como funci\'on de la energ\'ia del laser incidente, encontramos valores muy peque\~nos de las intensidades cuando $h\nu_i<E_{\rm{gap}}$. Sin embargo, si $h\nu_i=E_{\rm{gap}}$ la amplitud de transici\'on Raman aumenta de manera abrupta. Este comportamiento lo denominamos la regla del salto para la intensidad Raman y es mostrado gr\'aficamente en la figura \ref{fig8_ssc_aal2005} para los modos colectivos monopolares en un campo magn\'etico externo $B=$ 4.5 T.
El comportamiento de las intensidades Raman de estos mismo modos, en geometr\'ia polarizada para diferentes valores de $h\nu_i$ y del campo magn\'etico externo $B$, es mostrado en la figura \ref{fig4_prbrc_aal2005}. La regla del salto de la intensidad Raman es apreciable en la figura \ref{fig4_prbrc_aal2005} (b) para el modo colectivo de esp\'in en todos los valores de campo excepto en los puntos $B=0$ (regla de selecci\'on), 2.5, 3.0 y 5.0 T donde observamos un colapso de este comportamiento. Un an\'alisis del mismo tipo al presentado en la figura \ref{fig4_prbrc_aal2005} se realiz\'o para las excitaciones uniparticulares de bajas energ\'ias obteni\'endose un resultado similar \cite{AAL_PRBRC_2005}.
La regla del salto es obedecida por todos los modos monopolares para valores peque\~nos de $B$. Aunque esta regla no es universal respecto al campo magn\'etico externo constituye una herramienta \'util para la identificar la naturaleza de las excitaciones electr\'onicas en puntos cu\'anticos.

\subsection{Espectros Raman en r\'egimen de resonancia extrema ($h\nu_i \approx E_{\rm{gap}}$)}

Algunas de las caracter\'isticas discutidas en la secci\'on anterior se manifiestan tambi\'en en este r\'egimen, sin embargo aparecen nuevos efectos en los espectros Raman caracter\'isticos de la espectroscop\'ia Raman en r\'egimen de excitaci\'on resonante. 

\subsubsection{Predominio de los picos Raman correspondientes a excitaciones uniparticulares}

\begin{figure}[ht]
\begin{center}
\includegraphics[width=1\linewidth,angle=-90]{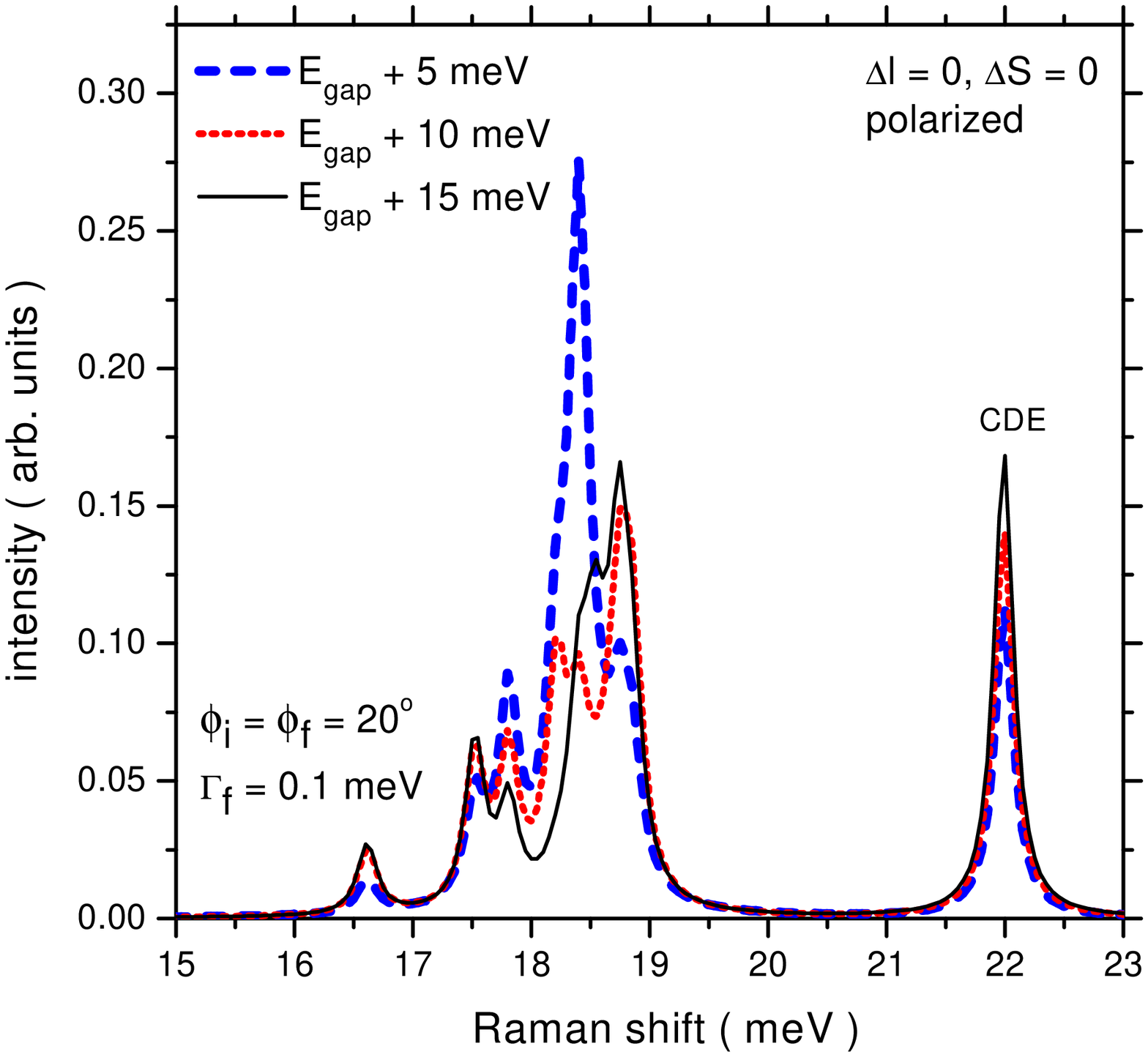}
\caption{\label{fig9_ssc_aal2005} Espectros Raman de excitaciones monopolares en r\'egimen de resonancia extrema. $B=0$ T.}
\end{center}
\end{figure}

Las intensidades Raman asociadas a las excitaciones uniparticulares experimentan un incremento notable en condiciones de excitaci\'on resonante. En este r\'egimen la mayor contribuci\'on en el c\'alculo de la amplitud de transici\'on Raman Ec. (\ref{eq1}) est\'a dada por el estado intermedio en resonancia con la energ\'ia del laser incidente. Este estado (virtual) decae indiscriminadamente a estados excitados uniparticulares o colectivos. Como fue mostrado en las figuras \ref{fig2_prb_aal2004} y \ref{fig1_prbrc_aal2005} existe un n\'umero grande de estados finales uniparticulares compactados en regiones estrechas de la energ\'ia de excitaci\'on (picos en la densidad de estados finales) contribuyendo esto a que los picos Raman en esta regi\'on sean mayores que aquellos correspondientes a los modos colectivos. Este resultado est\'a ilustrado en la figura \ref{fig9_ssc_aal2005} la cual muestra los espectros Raman monopolares para diferentes valores de $h\nu_i$.

\subsubsection{Correlaci\'on entre las intensidades de picos Raman individuales y la densidad de niveles de energ\'ia de los estados intermedios}

\begin{figure}[ht]
\begin{center}
\includegraphics[width=1\linewidth,angle=0]{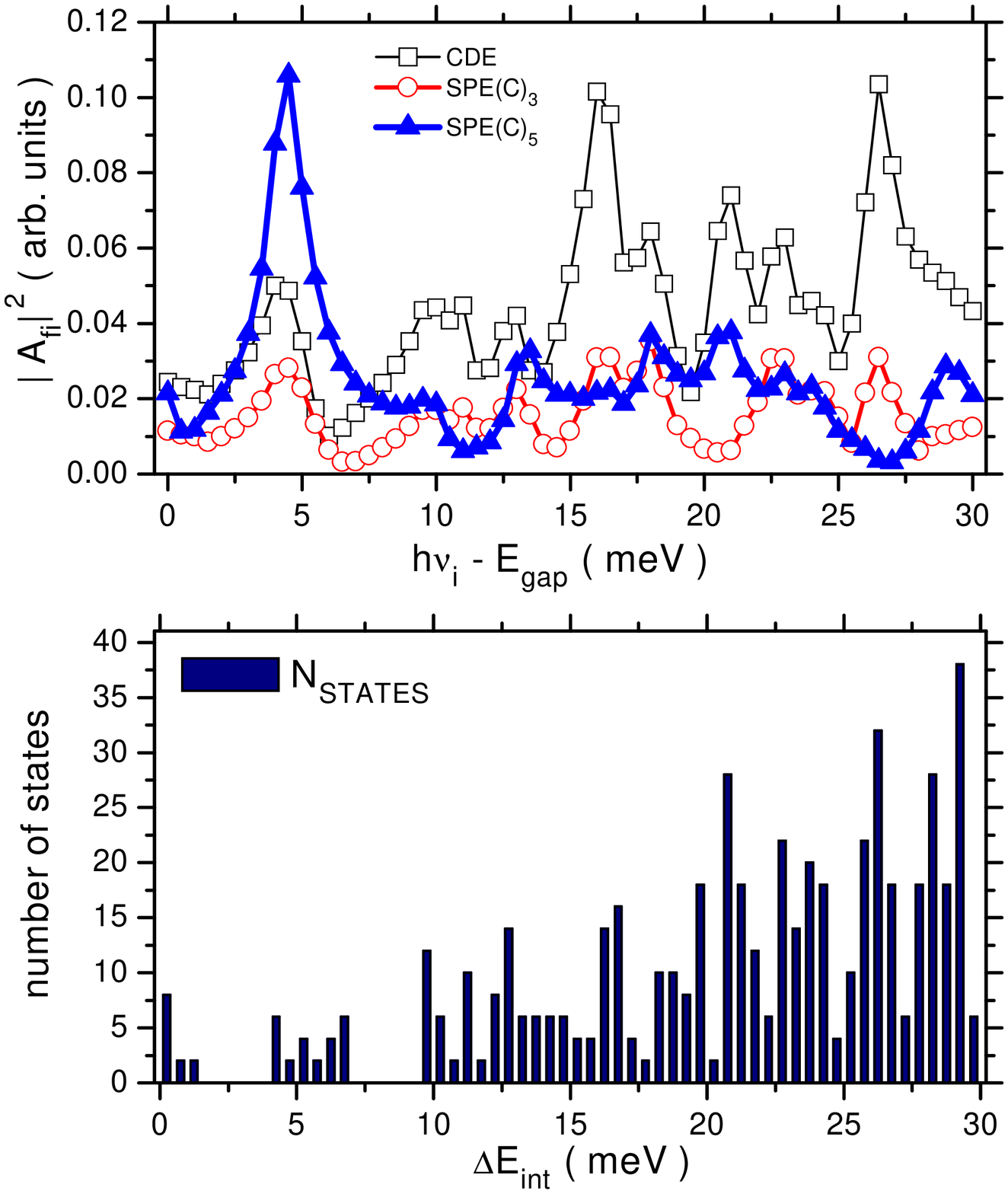}
\caption{\label{fig10_ssc_aal2005} Comportamiento de $\vert A_{\rm{fi}} \vert ^2$ asociados a excitaciones monopolares en r\'egimen resonante en conjunto con la densidad de estados intermedios. $B=0$ T.}
\end{center}
\end{figure}

En la figura \ref{fig10_ssc_aal2005} ploteamos los valores de $\vert A_{\rm{fi}} \vert ^2$ como funci\'on de $h\nu_i$ en el intervalo ($E_{\rm{gap}}, E_{\rm{gap}}+30$ meV) en conjunto con la densidad de niveles de energ\'ias de los estados intermedios para los mismos estados finales usados en la construcci\'on de la figura \ref{fig5_ssc_aal2005}. Como resultado de las resonancias con los estados intermedios podemos observar fluctuaciones abruptas en los valores de las amplitudes de transici\'on Raman ploteados. Los picos en los valores de $\vert A_{\rm{fi}} \vert ^2$ reproducen, fundamentalmente en la regi\'on donde $h\nu_i - E_{\rm{gap}} \leq 15$ meV, los inicios de los conglomerados de estados intermedios.   

\subsubsection{Efectos de interferencia en el c\'alculo de la amplitud de transici\'on Raman} 
\begin{figure}[ht]
\begin{center}
\includegraphics[width=1\linewidth,angle=-90]{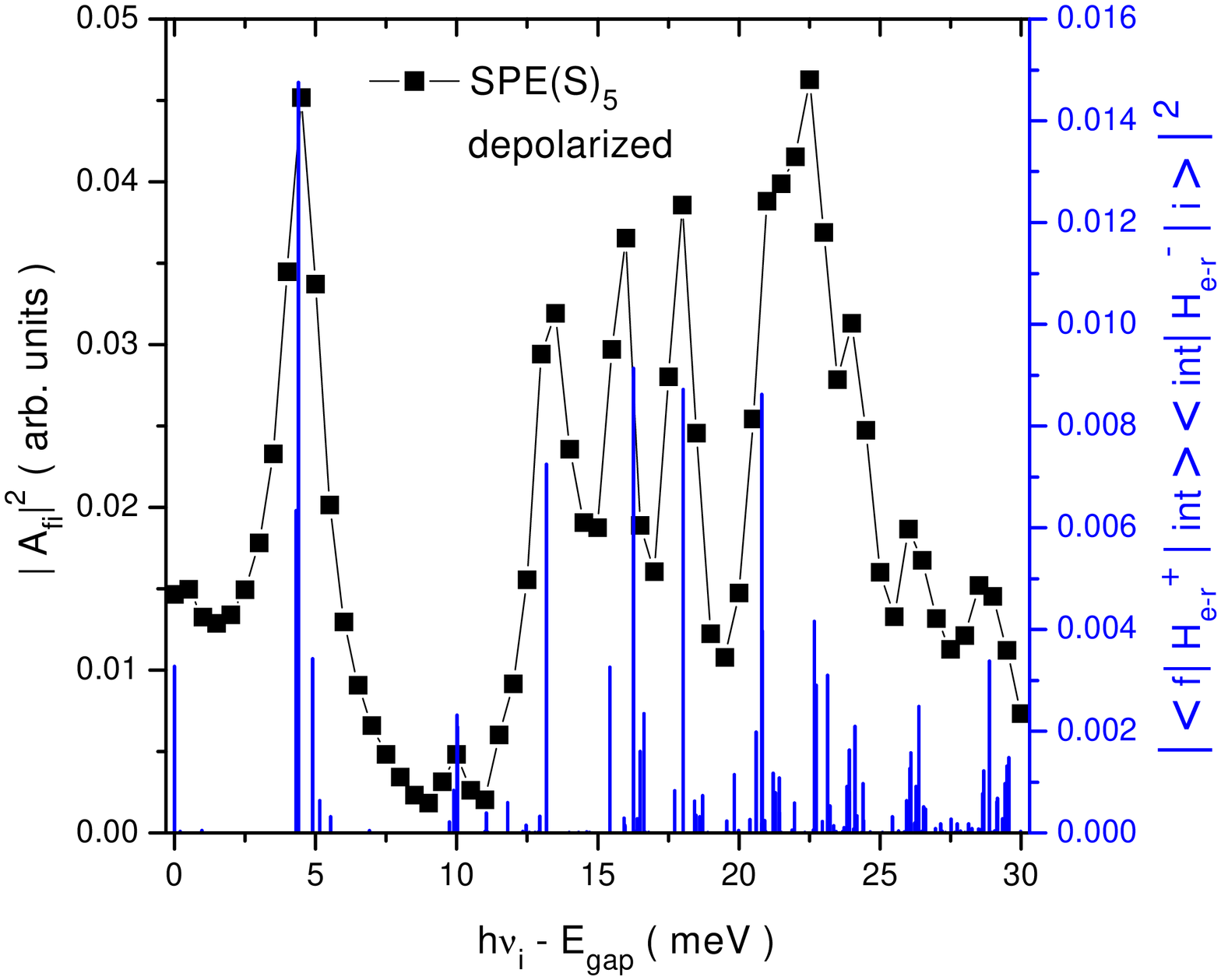}
\caption{\label{fig11_ssc_aal2005} $\vert A_{\rm{fi}} \vert ^2$ asociado a un estado excitado  uniparticular de esp\'in en conjunto con la contribuci\'on individual de cada estado intermedio en resonancia con $h\nu_i$. $B=0$ T.}
\end{center}
\end{figure}

En la figura \ref{fig11_ssc_aal2005} evaluamos la contribuci\'on de los diferentes estados intermedios que entran en la suma de la Ec. (\ref{eq1}). En esta figura comparamos la magnitud $\vert A_{\rm{fi}} \vert ^2$ asociada a un estado final (uniparticular) de esp\'in con las contribuciones individuales del estado intermedio en resonancia con la energ\'ia del laser incidente. Como conclusi\'on podemos extraer que s\'olo aquellos estados intermedios cuyas energ\'ias est\'an muy cercanas al valor de $h\nu_i$ contribuir\'an al valor de la amplitud de transici\'on Raman, por tanto, los efectos de inerferencia en el c\'alculo de $A_{\rm{fi}}$ son d\'ebiles.

\subsection{Espectros Raman en r\'egimen resonante en la regi\'on donde $h\nu_i \gg E_{\rm{gap}}$}

\begin{figure}[ht]
\begin{center}
\includegraphics[width=0.7\linewidth,angle=-90]{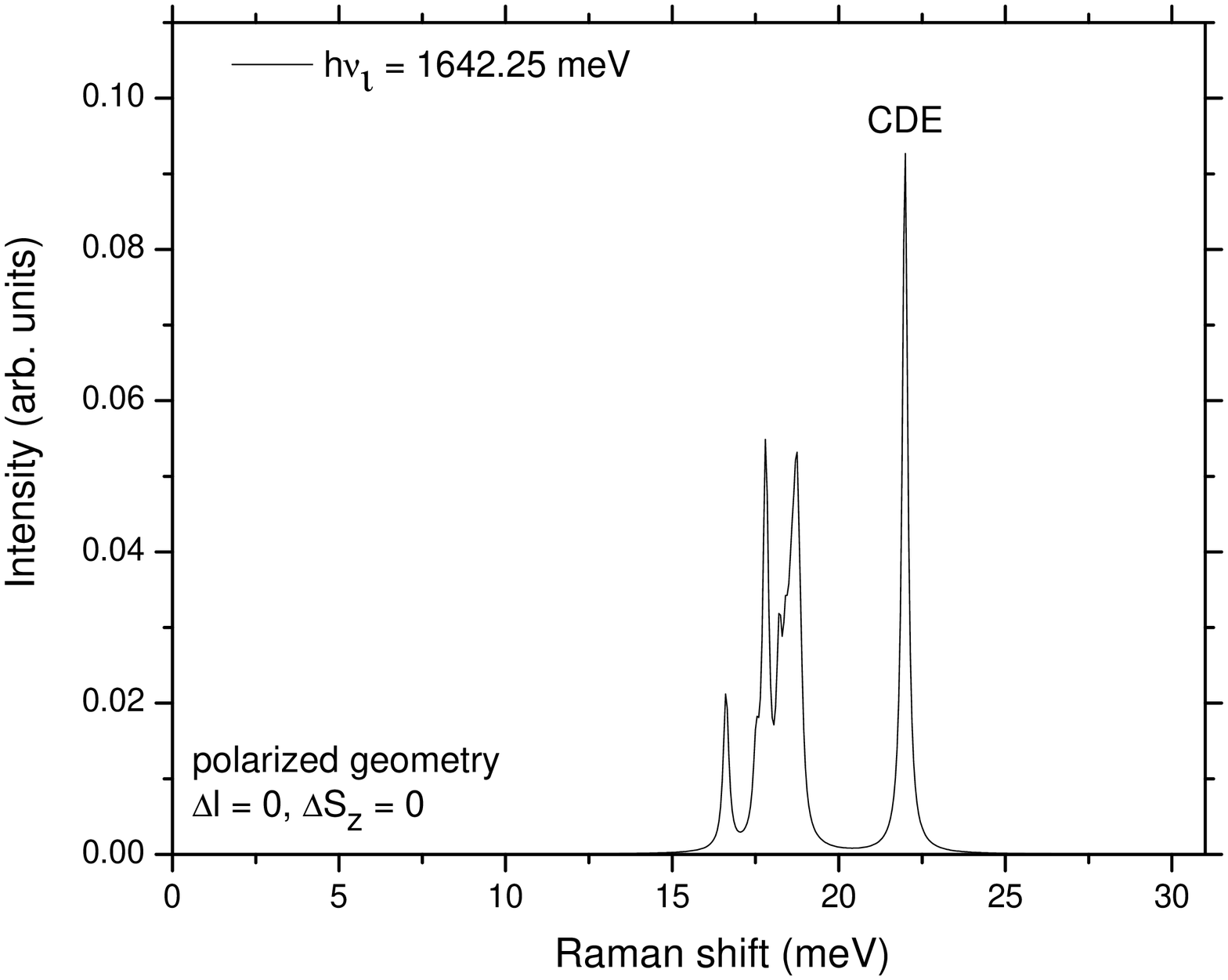}
\caption{\label{fig8_prb_aal2004} Espectro Raman monopolar en geometr\'ia polarizada. $h\nu_i=E_{\rm{gap}}+40$ meV. $B=0$ T.}
\end{center}
\end{figure}

Como mencionamos al inicio de la secci\'on (\ref{results}) para el c\'alculo de la amplitud de transici\'on Raman en este r\'egimen asumimos que el par\'ametro $\Gamma_{\rm{int}}$, asociado al tiempo de vida del estado intermedio $\vert \rm{int} \rangle$, experimenta un incremento abrupto cuando $E_{\rm{int}} > E_{\rm{gap}} + \hbar \omega_{\rm{LO}}$. Como resultado la contribuci\'on de estos estados a $A_{\rm{fi}}$ no es resonante, por lo que la intensidad de los picos Raman ser\'an funciones suaves respecto a $h\nu_i$ en similitud a lo explicado en la secci\'on \ref{smooth_bvelow_gap}. Sin embargo, un incremento en la intensidad Raman dado un valor particular de $h\nu_i$ en este r\'egimen no s\'olo puede estar dado por un incremento en el valor del producto $\langle f|\hat H^+_{e-r}|X\rangle\langle X|\hat H^-_{e-r}|i\rangle$, una disminuci\'on en el valor del par\'ametro $\Gamma_{\rm{int}}$ se manifiesta tambi\'en en un aumento de la intensidad Raman. En el conjunto de estados intermedios incluidos en el c\'alculo de la amplitud de transici\'on Raman pueden existir estados con propiedades f\'isicas especificas \cite{AAL_PRB_2004} que se manifiestan en una disminuci\'on del valor del par\'ametro $\Gamma_{\rm{int}}$ asociado a dichos estados. En la figura \ref{fig8_prb_aal2004} mostramos un espectro Raman calculado con $h\nu_i=E_{\rm{gap}}+40$ meV.
Esta energ\'ia de incidencia la hemos escogido en resonancia con un estado intermedio cuya anchura energ\'etica ha sido fijada a $\Gamma_{\rm{int}}=2$ meV. En la figura \ref{fig8_prb_aal2004} se puede observar un espectro Raman donde el pico mas intenso corresponde al modo colectivo monopolar de carga. La inclusi\'on de los tiempos de vida de los estados intermedios \cite{en_curso} en el esquema de c\'alculo implementado podr\'ia complementar cuantitativamente la idea cualitativa expuesta en esta secci\'on, lo cual explicar\'ia la fenomenolog\'ia experimental observada en este r\'egimen.      

\section{Conclusiones}   
\label{conclusions}
En este trabajo hemos implementado una esquema de c\'alculo consistente de la amplitud de transici\'on Raman asociada a excitaciones electr\'onicas en un punto cu\'antico cargado con 42 electrones. Los espectros Raman calculados han revelado caracter\'isticas muy interesantes y una gran sensibilidad con respecto a la energ\'ia de la luz incidente y a la presencia o no de campos magn\'eticos externos. En la regi\'on de r\'egimen no resonante, las intensidades de los picos Raman dependen suavemente de la energ\'ia del laser incidente. Para $B=0$, las reglas de selecci\'on del efecto Raman, deducibles a partir de la ORA para los modos colectivos, es obedecida tambi\'en por los estados excitados uniparticulares. Esto implica, que es posible obtener informaci\'on del perfil de la densidad de estados finales excitados a partir de la estructura del espectro Raman medido o calculado. En presencia de campos magn\'eticos externos, se rompen las reglas de selecci\'on de la polarizaci\'on del efecto Raman, particularmente cuando $h\nu_i$ es incrementado hasta la brecha energ\'etica del semiconductor, efecto que hemos denominado regla del salto de la intensidad Raman. En r\'egimen de excitaci\'on resonante las intensidades Raman muestran un comportamiento de fuertes fluctuaciones con la energ\'ia de la luz incidente como resultados de las resonancias con los estados intermedios. En este r\'egimen los picos mas intensos en los espectros est\'an en correspondencia con estados excitados uniparticulares. Se determin\'o que los efectos de interferencia en el c\'alculo de la amplitud de transici\'on Raman son d\'ebiles, quedando el valor de esta magnitud determinada por la contribuci\'on de los estados intermedios cuasi-resonantes con la energ\'ia del laser incidente. Una interpretaci\'on cualitativa para los espectros Raman en la regi\'on donde la energ\'ia de la radiaci\'on incidente est\'a muy por encima de la brecha energ\'etica del semiconductor es propuesta. Estos resultados complementan las interpretaciones que se han brindado a los trabajos experimentales publicados en el tema y estimulan la realizaci\'on de futuros experimentos que podr\'ian ayudar a la comprensi\'on de aspectos complejos de la espectroscop\'ia Raman de excitaciones electr\'onicas en puntos cu\'anticos.

\end{document}